\documentclass[review]{elsarticle}
\usepackage[colorlinks=true]{hyperref}
\usepackage{lineno,hyperref}
\usepackage{color}
\usepackage{amsmath,amsfonts,amssymb,esvect,ulem}
\usepackage{float}
\usepackage{graphicx}
\usepackage{epsfig}
\usepackage{cancel}
\usepackage{epstopdf}
\usepackage{subcaption}
\modulolinenumbers[5]
\newcommand{\sti}{\sigma_\mm{ti}}
\journal{Nuclear Science and Engineering}
\newcommand{\dnn}{\mathcal{D}_\mm{n}}
\newcommand{\dvv}{\mathcal{D}_\mm{v}}

\newcommand{\nno}{\nonumber}

\newcommand{\st}{\sigma_\mathrm{t}}

\newcommand{\psii}[1]{\psi^\mathrm{inc}_\mathrm{{#1}}}

\newcommand{\sn}{S$_N$}
\newcommand{\pn}{P$_N$}
\newcommand{\omen}{\ome\cdot\nabla}

\newcommand{\ve}[1]{\vec{{#1}}}
\newcommand{\mm}[1]{\mathrm{#1}}

\newcommand{\pd}{{\partial\mathcal{D}}}

\newcommand{\intli}[1]{\int\limits_{{#1}}}
\newcommand{\ointli}[1]{\oint\limits_{{#1}}}

\newcommand{\dii}{{\mathcal{D}_\mathrm{i}}}
\newcommand{\djj}{{\mathcal{D}_\mathrm{j}}}
\newcommand{\fij}{{\mathcal{F}_\mathrm{i,j}}}
\newcommand{\pdi}{{\partial\mathcal{D}_\mathrm{i}}}

\newcommand{\slim}[1]{\sum\limits_{\tiny#1}}
\newcommand{\dint}{\oint\limits_{4\pi}d\Omega\ \int\limits_\mathcal{D}dV\ }
\newcommand{\nido}{\vec{n}_\mathrm{i}\cdot\ome}
\newcommand{\absnido}{\left|\nido\right|}
\newcommand{\ndo}{\vec{n}\cdot\ome}
\newcommand{\absndo}{\left|\ndo\right|}

\newcommand{\e}[1]{\ensuremath{\times 10^{#1}}}
\newcommand{\TAMU}{Texas A\&M University}

\newcommand{\sigs}{\sigma_\mathrm{s}}
\newcommand{\siga}{\sigma_\mathrm{a}}

\newcommand{\qs}{q_\mathrm{s}}
\newcommand{\ome}{\vv{\Omega}}
\newcommand{\dome}{d\Omega}

\newcommand{\jo}{j^\mathrm{out}}

\newcommand{\lp}{\left(}
\newcommand{\rp}{\right)}

\newcommand{\red}[1]{\textcolor{black}{#1}}
\begin{document}

\begin{frontmatter}

\title{An Accurate Globally Conservative Subdomain Discontinuous Least-squares Scheme for Solving Neutron Transport Problems}

\author[mycurrentaddress]{Weixiong Zheng\corref{mycorrespondingauthor}}
\cortext[mycorrespondingauthor]{Corresponding author}
\ead{weixiong.zheng.uc.berkeley@gmail.com}

\author[mymainaddress]{Ryan G. McClarren}
\ead{rgm@tamu.edu}

\author[mymainaddress]{Jim E. Morel}
\ead{morel@tamu.edu}
\address[mycurrentaddress]{Nuclear Engineering, University of California,\ Berkeley,~Berkeley, CA 94709}
\address[mymainaddress]{Nuclear Engineering, \TAMU,~College Station, TX 77843-3133}

\begin{abstract}
	In this work, we present a subdomain discontinuous least-squares (SDLS) scheme for neutronics problems. {Least-squares (LS) methods are known to be inaccurate for problems with \red{sharp} total-cross section interfaces.} In addition, the least-squares scheme is known not to be globally conservative in heterogeneous problems. In problems where global conservation is important, e.g. k-eigenvalue problems, a conservative treatment must be applied. We, in this study, propose an SDLS method that retains global conservation, and, as a result, gives high accuracy on eigenvalue problems. Such a method resembles the LS formulation in each subdomain without a material interface and differs from LS {in}\ that an additional least-squares interface term appears for each interface. The scalar flux is continuous in each subdomain with continuous finite element method (CFEM) while discontinuous on interfaces for every pair of contiguous subdomains. \red{SDLS numerical results are compared with those obtained from other numerical methods with test problems having material interfaces. High accuracy of scalar flux in fixed-source problems and $k_\mathrm{eff}$\ in eigenvalue problems are demonstrated.}
\end{abstract}


\end{frontmatter}


\section{Introduction}
Neutral particle transport problems are governed by a first-order, hyperbolic equation.  As a result particle transport problems, especially those that are streaming dominated, can have {sharp gradients along the characteristic lines}. Such problems require spatial discretizations that allow for discontinuities in space.  Moreover, in regions where there is a strong scattering, discontinuous finite element method (DFEM) have been shown to properly preserve the asymptotic diffusion limit of the transport equation \cite{adams2001discontinuous}.  These two facts have led to discontinuous Galerkin finite elements being widely {accepted} in transport calculations. Nevertheless, the discontinuous finite element methods have more degrees of freedom {(DoFs)}\ than their continuous counterparts, especially in 3-D. 

Another approach to solving transport problems involves forming a second-order transport operator, and solving the resulting equations using continuous finite elements. Second-order transport problems based on the parity of the equations and the self-adjoint angular flux (SAAF) equation are well-known\cite{morel_saaf,gesh_dissertation,gesh_mc,yunhuang_dissertation,zheng_dissertation}. The resulting equations are symmetric, but are ill-posed in void regions and ill-conditioned in near voids.  

More recently, Hansen, et al.~\cite{Hansen:2015jq} derived a second-order form that is equivalent to minimizing the squared residual of the transport operator; it is therefore called the least-squares transport equation (LSTE).  This method can also be formed by multiplying the transport equation by the adjoint transport operator or by applying least-squares finite elements in space to the transport equation.  The resulting equations are well-posed in void and are symmetric positive definite \red{(SPD)}. The least-squares method has been continuously investigated  in the {applied mathematics} and computational fluid dynamics communities {for decades}\ \cite{ls-60s,ls-70s,ls-80s,lowrie_l1,guermond_l1,bochev-ls}. However, the method generally has low accuracy in problems with interfaces between optically thin and thick materials without local refinement near the interface\cite{zheng_l1sn,zheng-l1,zheng_l1pn,clifmc}. Additionally, the standard least-squares method does not have particle conservation (except in the limit as the number of zones goes to infinity), which {would induce undesirably large $k$-eigenvalue errors in neutronics simulations} \cite{laboure:2016}, unless {conservative acceleration schemes, such as nonlinear diffusion acceleration described in \cite{morel_holo,park-nda,yaqi_void},}\ are applied to {regain the conservation}.

In the process of deriving the least-squares equations, the symmetrization of the transport equation converts a hyperbolic equation to an elliptic equation and \red{causes loss of causality}. For example, the presence of a strong absorber downstream of a source can influence the solution upstream of the source. In remedying this deficiency, {it is desirable} to add back some asymmetry to the operator, but do so in such a way that still preserves the positive properties of the least-squares method.

To this end we develop a least-squares method that minimizes the square of the transport residual over certain regions of the entire problem domain. In each of these subregions the interaction cross-sections are slowly varying functions of space. We allow the solution between these regions to be discontinuous. The resulting scheme essentially solves least-squares independently in each subregion, and the regions are coupled through a sweep-like procedure\footnote{{In this work, we will utilize the discrete-ordinates method for angular discretization. When solving the discrete-ordinates equations with first-order discretization techniques, such as discontinuous Galerkin method, the procedure is such that the transport equation is solved from upstream cells to downstream cells, a procedure called a ``sweep".}}. Furthermore, with the correct weighting in the least-squares procedure, we can construct a method that is globally conservative.

The reminder of the paper is organized as follows: {we start off reviewing the governing equation and the ordinary least-squares finite element weak formulation derived from the minimization point of view in Sec.\ \ref{s:ls}}.\ In Sec.\ \ref{s:cdls_derive},\ we propose a new method that is based on a least-squares formulation over contiguous blocks in the problem base on a novel subdomain-discontinuous functional. We also derive the corresponding weak formulations in this section. Next, we further theoretically demonstrate the method, unlike standard least-squares methods, retains conservation in both a global and subdomain sense. In Sec.\ \ref{s:numerics}, several numerical tests are presented to demonstrate conservation as well as the improved accuracy. {We then conclude the study and discuss potential future work in Sec.\ \ref{s:conc}.}

\section{Revisit least-squares discretization of transport equation}\label{s:ls}
\subsection{One group transport equation}
We will consider steady, energy-independent transport problems with isotropic scattering in this work.  The complications of energy dependence and anisotropic scattering can be readily incorporated into our method. The steady transport equation used to describe neutral particles of a single speed is expressed in operator form as:
\begin{subequations}\label{e:te}
\begin{equation}
L\psi=\qs,
\end{equation}
where {the transport operator $L$\ is defined as the sum of the streaming operator $\omen(\cdot)$\ and total collision operator $\st$:}
\begin{equation}
L=\omen(\cdot)+\st,
\end{equation}
{and $\qs$\ represents the total volumetric source defined as the sum of the scattering  source $S\psi$\ and a fixed volumetric source $q(\ve{r},\ome)$:}
\begin{equation}
\qs=S\psi+q.
\end{equation}
\end{subequations}
In these equations $\psi(\ve{r},\ome)$ is the angular flux of neutral particles with units of particles per unit area per unit time, where $\ve{r} \in {\mathcal{D}\subset}\mathbb{R}^3$,\ {$\mathcal{D}$\ stands for problem domain}\ and $\ome \in \mathbb{S}_2$ is a point on the unit sphere representing a direction of travel for the particles. $S$ is the scattering operator. For isotropic scattering, it is defined as 
\begin{align}
{S \psi = \int\limits_{4\pi}\dome'\ \sigs\left(\ome\cdot\ome'\right)\psi(\ve{r},\ome')=\frac{\sigs\phi}{4\pi},}
\end{align}  
\red{where $\sigma_\mm{s}\left(\ome\cdot\ome'\right)$\ is differential scattering cross section defined as}
\begin{align}
\red{\sigma_\mm{s}\left(\ome\cdot\ome'\right)=\frac{\sigma_\mm{s}}{4\pi}}
\end{align}
and $\phi$\ is the scalar flux defined as
\begin{align}
{\phi=\intli{4\pi}\dome\ \psi.}
\end{align}
Additionally, $\sigs$\ {is the scattering cross section.}

The boundary conditions for Eq.~\eqref{e:te} specify the angular flux $\psii{}$ on the boundary for incoming directions:
\begin{equation}\label{e:bc}
\psi(\ve{r}, \ome) = \psii{}(\ve{r}, \ome) \qquad \text{for}\quad r\in\pd, \qquad \ndo<0.
\end{equation}

{We discretize the angular component of the transport equation using discrete-ordinates (\sn)\ method \cite{glasstone}.\ Therein, we use a quadrature set \{$w_m,{\ome}_m$\},\ containing weights $w_m$\ and quadrature collocation points ${\ome}_m$,\ for the angular space to obtain the set of equations:
	\begin{align}
	{\ome}_m\cdot\nabla\psi_m+\st\psi_m=q_\mm{s},\quad\red{\psi_m=\psi\left(\ome_m\right)},\quad m=1,\cdots,M,
	\end{align}
	with $M$\ being the total number of angles in the quadrature. Additionally, the angular integration for generic function $f$\ is defined as:
	\begin{align}
	\intli{4\pi}\dome\ f(\ome)=\slim{m=1}^Mw_mf(\ome_m).
	\end{align}
	For instance, the scalar flux is expressed as
	\begin{align}
	\phi(\vec{r})=\intli{4\pi}\dome\ \psi(\vec{r},\ome)\approx\slim{m=1}^Mw_m\red{\psi_m(\vec{r})}.
	\end{align}}

\subsection{LS weak formulation}
In order to employ CFEM to solve the transport equation, Hansel, et al.\ \cite{Hansen:2015jq} derived a least-squares form of transport equation by multiplying the transport equation by adjoint streaming and removal operator,\ i.e.
\begin{equation}\label{eq:ls_adj}
L^\dagger\left(L\psi-\qs\right)=0,
\end{equation}
where
\begin{align}
{L^\dagger = - \omen(\cdot)+\st.}
\end{align}
The corresponding weak form of the problem in  Eq.~\eqref{eq:ls_adj} with CFEM  is: 
\red{given a function space $\mathcal{V}$},\ find $\psi\in\mathcal{V}$\ such that $\forall v\in\mathcal{V}$:
\begin{equation}\label{e:hansen1}
\dint vL^\dagger\left(L\psi-\qs\right)=0.
\end{equation}
{The property of adjoint operators allows us to express Eq.~\eqref{e:hansen1} as
\begin{equation}\label{eq:weakform_final0}
\dint Lv\left(L\psi-\qs\right)-\ointli{4\pi}\dome\ \intli{\pd}ds\ {\ndo v(L\psi-\qs)}=0.
\end{equation}
Additionally, we require the transport equation to be satisfied on the boundary as well:
\begin{equation}\label{eq:bc0}
L\psi-\qs=0.
\end{equation}
Therefore, the weak form can be expressed as:
\begin{equation}\label{eq:weakform_final}
\dint Lv\left(L\psi-\qs\right)=0.
\end{equation}}

On the other hand, one could also derive the weak form from defining a least-squares functional of transport residual:
\begin{equation}\label{e:func0}
\Gamma_\mathrm{LS}=\frac{1}{2}\oint\limits_{4\pi}d\Omega\ \int\limits_\mathcal{D}dV\ (L\psi-\qs)^2.
\end{equation}
Minimizing Eq.\ \eqref{e:func0}\ in a discrete function space $\mathcal{V}$\ leads to Eq.~\eqref{eq:weakform_final}\ as well\footnote{The procedure of the minimization is to find a stationary ``point" in {a}\ specific function space, see Ref.\ \cite{runchang}\ for details.}.

\subsection{Imposing boundary {conditions}}\label{s:ls-cons}
In the derivation above, the incident boundary conditions {have been ignored}. Though it is possible to impose a strong boundary condition \cite{yunhuang_dissertation},\ a weak boundary condition is chosen in this work instead. To weakly impose the boundary condition, the functional is changed to
\begin{align}
\Gamma_\mathrm{LS}=\frac{1}{2}\oint\limits_{4\pi}d\Omega\ \int\limits_\mathcal{D}dV\ (L\psi-\qs)^2+\frac{1}{2}\int\limits_{\ndo<0}d\Omega\ \int\limits_\mathcal{\pd}ds\ \lambda \lp\psi-\psii{}\rp^2, 
\end{align}
where $\lambda$ is {a freely chosen Lagrange multiplier}. The resulting LS weak formulation with boundary condition is then expressed as:
\begin{align}\label{e:ls1}
\oint\limits_{4\pi}d\Omega\ \int\limits_\mathcal{D}dV\ Lv\left(L-S\right)\psi+\int\limits_{\ndo<0}d\Omega\ \int\limits_\mathcal{\pd}ds\ \lambda v\psi\nonumber\\
=\oint\limits_{4\pi}d\Omega\ \int\limits_\mathcal{D}dV\ Lvq+\int\limits_{\ndo<0}d\Omega\ \int\limits_\mathcal{\pd}ds\ \lambda v\psii{},
\end{align}
where the weight function $v$\ is any element of the trial space.

If we choose $\lambda = \st\absndo$, then in non-void situations the LS scheme is globally conservative in homogeneous media (i.e., where $\st$\ is spatially independent in the whole domain). This can be seen by taking the weight function $v$\ to be $1$\ in Eq.\ \eqref{e:ls1} and performing the integration over angle to get
\begin{subequations}\
\begin{align}
\st\mathbf{B}=0,
\end{align}
\begin{equation}\label{e:balance0}
\mathbf{B}:=\lp\intli{\pd}ds\ \jo-\intli{\ndo<0}\dome \intli{\pd}ds\ \absndo\psii{}+\intli{\mathcal{D}}dV(\siga\phi-Q)\rp,
\end{equation}
\begin{equation}
\jo:=\intli{\ndo>0}\dome\ \absndo\psi\quad\mathrm{and}\quad Q:=\intli{4\pi}\dome\ q.
\end{equation}
\end{subequations}
Here $\mathbf{B}$ is the global balance: it states that the outgoing current, $\jo$, plus the absorption $\siga\phi$ is equal to the total source plus the incoming current. 
{When $\st=0$,\ $\mathbf{B}$\ being any value does not disturb $\st\mathbf{B}=0$.\ In general, what we observe is $\mathbf{B}\neq0$.\ Therefore, conservation is lost.}

\section{A least-squares discretization allowing discontinuity on subdomain interface}\label{s:cdls_derive}

\subsection{The SDLS functional and weak formulation}

Given that LS is conservative  when the cross-sections are constant and non-void, we propose to reformulate the problem to solve  using least-squares in subdomains where cross-sections are constant. On the boundary of these subdomains we connect the angular fluxes using an interface condition that allows the fluxes to be discontinuous. The result is a discretization that can be solved using a procedure analogous to transport sweeps where the sweeps are over subdomains, instead of mesh zones.

In the context of a minimization problem, we can define a functional in the following form:
\begin{align}\label{e:cdls_func}
\Gamma_\mathrm{SDLS}=\frac{1}{2}\slim{\dii}&\ointli{4\pi}d\Omega\ \intli{\dii}dV\ (L_\mathrm{i}\psi_\mathrm{i}-q_\mathrm{si})^2+{\frac{1}{2}\slim{\dii\cap\pd\neq\emptyset}\intli{\nido<0}d\Omega\ \intli{\pd\cap\dii}ds\ \sigma_\mm{ti}\absnido(\psi_\mm{i}-\psii{})^2}\nonumber\\
&+\frac{1}{2}\slim{\dii}\slim{\fij}\intli{\nido<0}d\Omega\ \intli{\fij}ds\ \sigma_\mathrm{ti}\absnido({\psi_\mm{i}-\psi_\mm{j}})^2,
\end{align}
where $\fij$\ is the interface between $\dii$\ and any contiguous subdomain $\djj$.\ Accordingly, the variational problem turns to: find $\psi_\mathrm{i}$\ in a polynomial space $\mathcal{V}$\ such that $\forall v_\mathrm{i}\in\mathcal{V}$,\
\begin{align}\label{e:cdls_weak}
&\slim{\dii}\left(\ointli{4\pi}d\Omega\ \intli{\dii}dV\ L_\mathrm{i}v_\mathrm{i}\left(L_\mathrm{i}-S_\mathrm{i}\right)\psi_\mathrm{i}+\slim{\fij}\sigma_\mathrm{ti}\intli{\nido<0}d\Omega\ \intli{\fij}ds\  \absnido v_\mathrm{i}\left(\psi_\mathrm{i}-\psi_\mathrm{j}\right)\right)\nno\\
&+{\slim{\dii\cap\pd\neq\emptyset}\intli{\nido<0}d\Omega\ \intli{\pd\cap\dii}ds\ \sigma_\mm{ti}\absnido v_\mm{i}\psi_\mm{i}}=\slim{\dii}\ointli{4\pi}d\Omega\ \intli{\dii}dV L_\mathrm{i}v_\mathrm{i}q_\mathrm{i}\\
&+{\slim{\dii\cap\pd\neq\emptyset}\intli{\nido<0}d\Omega\ \intli{\pd\cap\dii}ds\ \sigma_\mm{ti}\absnido v_\mm{i}\psii{}.}\nno
\end{align}

Compared with ordinary LS method as illustrated in Eq.\ \eqref{e:ls1},\ SDLS does not enforce the continuity on subdomain interface. That presents possibility of combining the least-squares method and transport sweeps. For a given direction, $\ome$, Eq.~\eqref{e:cdls_weak}\ can be written as a block-lower triangular system {if no re-entering interface manifests}. Therein, each block is LS applied to a subdomain.The inversion of this system requires solving a LS system for each subdomains, connected via boundary angular fluxes. 


\subsection{Subdomain-wise and global conservation}
{A favorable SDLS property is both subdomain conservation and global conservation are preserved. The demonstration is similar to the derivation in Sec.\ \ref{s:ls-cons}.\ Taking $v_\mm{i}=1$,\ $L_\mm{i}v_\mm{i}$\ simplifies to
\begin{equation}\label{e:test}
L_\mm{i}v_\mm{i}=\sti.
\end{equation}
Accordingly, the SDLS weak form in Eq.\ \eqref{e:cdls_weak}\ is transformed to
\begin{align}\label{e:cdls-balance0}
&\slim{\dii}\left(\ointli{4\pi}d\Omega\ \intli{\dii}dV\ \sti\left(L_\mathrm{i}-S_\mathrm{i}\right)\psi_\mathrm{i}+\slim{\fij}\intli{\nido<0}d\Omega\ \intli{\fij}ds\  \absnido\sti\left(\psi_\mathrm{i}-\psi_\mathrm{j}\right)\right)\nno\\
&+{\slim{\dii\cap\pd\neq\emptyset}\intli{\nido<0}d\Omega\ \intli{\pd\cap\dii}ds\ \sigma_\mm{ti}\absnido\psi_\mm{i}}=\slim{\dii}\ointli{4\pi}d\Omega\ \intli{\dii}dV \sti q_\mathrm{i}\\
&+{\slim{\dii\cap\pd\neq\emptyset}\intli{\nido<0}d\Omega\ \intli{\pd\cap\dii}ds\ \sigma_\mm{ti}\absnido\psii{}.}\nno
\end{align}
Denote the boundary of subdomain $\dii$\ by $\pdi$\footnote{{Note that $\partial\dii$\ could either be on the problem boundary or interior interfaces.}},\ the first integral in Eq.\ \eqref{e:cdls-balance0}\ can be expressed as:
\begin{align}\label{e:cdls-part}
&\slim{\dii}\ointli{4\pi}d\Omega\ \intli{\dii}dV\ \sti\left(L_\mathrm{i}-S_\mathrm{i}\right)\psi_\mathrm{i}=-\slim{\dii}\intli{\pdi}ds\ \intli{\nido<0}\dome\ \absnido\sti\psi_\mm{i}\nno\\
&+\slim{\dii}\intli{\pdi}ds\ \intli{\nido>0}\dome\ \absnido\sti\psi_\mm{i}+\slim{\dii}\intli{\dii}dV\ \sti\sigma_\mm{ai}\phi_\mm{i}.
\end{align}
Note that
\begin{align}\label{e:note}
&\slim{\dii}\slim{\fij}\intli{\nido<0}d\Omega\ \intli{\fij}ds\  \absnido\sti\psi_\mathrm{i}+{\slim{\dii\cap\pd\neq\emptyset}\intli{\nido<0}d\Omega\ \intli{\pd\cap\dii}ds\ \sigma_\mm{ti}\absnido\psi_\mm{i}}\nno\\
&-\slim{\dii}\intli{\pdi}ds\ \intli{\nido<0}\dome\ \absnido\sti\psi_\mm{i}=0.
\end{align}
Introducing Eqs.\ \eqref{e:cdls-part}\ and \eqref{e:note}\ into \eqref{e:cdls-balance0}\ leads to
\begin{align}
&\slim{\dii}\left(\intli{\dii}dV\ \sti\sigma_\mm{ai}\phi_\mm{i}+\slim{\pdi}\intli{\nido>0}d\Omega\ \intli{\fij}ds\  \sti\absnido\psi_\mathrm{i}\right)\nno\\
&=\slim{\dii}\slim{\fij}\intli{\nido<0}d\Omega\ \intli{\fij}ds\  \absnido\sti\psi_\mathrm{j}+\slim{\dii}\intli{\dii}dV \sti Q_\mathrm{i}\\
&+{\slim{\dii\cap\pd\neq\emptyset}\intli{\nido<0}d\Omega\ \ointli{\pd\cap\dii}ds\ \sigma_\mm{ti}\absnido\psii{}.}\nno
\end{align}
Define the incoming currents from contiguous subdomain $\djj$\ and problem boundary for subdomain $\dii$\ and outgoing current as
\begin{subequations}
\begin{align}
j_\mm{i,j}^\mm{in}(\vec{r}):=\intli{\nido<0}\dome\ \absnido\psi_\mm{i},
\end{align}
~
\begin{align}
j_\mm{i,{\bf b}}^\mm{in}(\vec{r}):=\intli{\nido<0}\dome\ \absnido\psii{}
\end{align}
and
\begin{align}
j_\mm{i}^\mm{out}(\vec{r}):=\intli{\nido>0}\dome\ \absnido\psi_\mm{i},
\end{align}
\end{subequations}
respectively. It is then straightforward to get
\begin{align}\label{e:cdls-balance1}
&\slim{\dii}\left(\intli{\dii}dV\ \sti\sigma_\mm{ai}\phi_\mm{i}+\intli{\pdi}ds\  \sti j_\mm{i}^\mm{out}\right)\nno\\
&=\slim{\dii}\slim{\fij}\intli{\fij}ds\  \sti j_\mm{i,j}^\mm{in}+{\slim{\dii\cap\pd\neq\emptyset}\intli{\pd\cap\dii}ds\ \sigma_\mm{ti}j_\mm{i,{\bf b}}^\mm{in}}+\slim{\dii}\intli{\dii}dV \sti Q_\mathrm{i}.
\end{align}
Assuming $\sti$\ is spatially uniform within $\dii$,\ we can further transform Eq.\ \eqref{e:cdls-balance1}\ to
\begin{align}\label{e:cdls-balance2}
&\slim{\dii}\sti\left(\intli{\dii}dV\ \sigma_\mm{ai}\phi_\mm{i}+\intli{\pdi}ds\ j_\mm{i}^\mm{out}\right)\nno\\
&-\slim{\dii}\sti\slim{\fij}\intli{\fij}ds\   j_\mm{i,j}^\mm{in}-{\slim{\dii\cap\pd\neq\emptyset}\sti\intli{\pd\cap\dii}ds\ j_\mm{i,{\bf b}}^\mm{in}}-\slim{\dii}\sti\intli{\dii}dV\ Q_\mathrm{i}=0.
\end{align}
or equivalently
\begin{align}\label{e:wtd_balance}
\slim{\dii}\sti{\bf B}_\mm{i}=0
\end{align}
For all nonzero $\sti$,\ in order to make Eq.\ \eqref{e:wtd_balance}\ true, one must have subdomain-wise conservation:
\begin{equation}\label{e:wted-conservation}
{\bf B}_\mm{i}\equiv0,\quad\forall\dii\subset\mathcal{D}
\end{equation}
Additionally, this implies that there is the global conservation:
\begin{align}
\slim{\dii}{\bf B}_\mm{i}=0
\end{align}
}

\subsection{Conservative void treatment}
{CFEM-SAAF is globally conservative, yet, not compatible with void and potentially ill-conditioned in near-void situations. Therefore, efforts have been put in alleviating CFEM-SAAF in void \cite{clifmc,laboure:2016,yaqi_void,vincent-physor16}.\ We will specially give a brief review of the treatment developed in \cite{laboure:2016,vincent-physor16}.
}

{
Therein, Laboure, et al.\ developed a hybrid method compatible with void and near-void based on SAAF and LS. Denote non-void and void/uniform near-void subdomains by $\dnn$\ and $\dvv$,\ respectively\footnote{{In our work, $\dvv$\ is defined with $\st<0.01$\ cm$^{-1}$.}}. Then the hybrid formulation is presented as:
\begin{align}\label{e:saaf-cls}
&\ointli{4\pi}\dome\ \intli{\mathcal{D}}dV\ \left(\tau\omen v\psi+\st v\psi-(1-\st\tau)\omen v\psi\right)+\intli{\ndo>0}\dome\ \intli{\pd}ds\ \absndo v\psi\nno\\
&=\ointli{4\pi}\dome\ \intli{\mathcal{D}}dV\ \left(\tau\omen v+v\right)\qs+\intli{\ndo<0}\dome\ \intli{\pd}ds\ \absndo v\psii{},
\end{align}
where
\begin{align}
\tau=\begin{cases}
1/\st,&\vec{r}\in\dnn\\
1/c&\vec{r}\in\dvv\\
\end{cases}.
\end{align}
$c$\ is a freely chosen constant and usually set to be $1$. One notices that in non-void subdomain, the formulation resembles CFEM-SAAF. At the same time,  global conservation is preserved. {{Therefore, the method is named ``SAAF-Conservative LS (SAAF-CLS)"}}.
}

{Consider a homogeneous near-void problem. The weak form in Eq.\ \eqref{e:saaf-cls}\ can be transformed to
\begin{align}\label{e:vt}
&\ointli{4\pi}\dome\ \intli{\dvv}dV\ Lv\left(L\psi-\qs\right)+\intli{\ndo<0}\dome\ \intli{\partial\dvv}ds\ \st\absndo v\left(\psi-\psii{}\right)\nno\\
&+\ointli{4\pi}\dome\ \intli{\dvv}dV\ cv\left(L\psi-\qs\right)+\intli{\ndo<0}\dome\ \intli{\partial\dvv}ds\ c\absndo v\left(\psi-\psii{}\right)=0.
\end{align}
In void/near-void, CLS is conservative:
\begin{align}
(\st+c){\bf B}=0.
\end{align}

We therefore modify SDLS such that void treatment based on Eq.\ \eqref{e:vt} is incorporated:
\begin{align}
&\slim{\dii\subset\dnn}\left(\ointli{4\pi}d\Omega\ \intli{\dii}dV\ L_\mathrm{i}v_\mathrm{i}\left(L_\mathrm{i}-S_\mathrm{i}\right)\psi_\mathrm{i}+\slim{\fij}\sigma_\mathrm{ti}\intli{\nido<0}d\Omega\ \intli{\fij}ds\  \absnido v_\mathrm{i}\left(\psi_\mathrm{i}-\psi_\mathrm{j}\right)\right)\nno\\
&+{\slim{\substack{\dii\cap\pd\neq\emptyset\\ \dii\subset\dnn}}\intli{\nido<0}d\Omega\ \intli{\pd\cap\dii}ds\ \sigma_\mm{ti}\absnido v_\mm{i}\psi_\mm{i}}\nno\\
&+\slim{\dii\subset\dvv}\left(\ointli{4\pi}d\Omega\ \intli{\dii}dV\ (cv_\mm{i}+L_\mathrm{i}v_\mathrm{i})\left(L_\mathrm{i}-S_\mathrm{i}\right)\psi_\mathrm{i}+\slim{\fij}\intli{\nido<0}d\Omega\ \intli{\fij}ds\ (c+\sti)\absnido v_\mathrm{i}\left(\psi_\mathrm{i}-\psi_\mathrm{j}\right)\right)\nno\\
&+{\slim{\substack{\dii\cap\pd\neq\emptyset\\ \dii\subset\dvv}}\intli{\nido<0}d\Omega\ \intli{\pd\cap\dii}ds\ (c+\sti)\absnido v_\mm{i}\psi_\mm{i}}\\
&=\slim{\dii\subset\dnn}\ointli{4\pi}d\Omega\ \intli{\dii}dV L_\mm{i}v_\mm{i}q_\mathrm{i}+{\slim{\substack{\dii\cap\pd\neq\emptyset\\ \dii\subset\dnn}}\intli{\nido<0}d\Omega\ \intli{\pd\cap\dii}ds\ \sti\absnido v_\mm{i}\psii{}.}\nno\\
&+\slim{{\dii\subset\dvv}}\ointli{4\pi}d\Omega\ \intli{\dii}dV (cv_\mm{i}+L_\mm{i}v_\mm{i})q_\mathrm{i}+{\slim{\substack{\dii\cap\pd\neq\emptyset\\ \dii\subset\dvv}}\intli{\nido<0}d\Omega\ \intli{\pd\cap\dii}ds\ (c+\sti)\absnido v_\mm{i}\psii{}.}\nno
\end{align}
}

\section{Numerical results}\label{s:numerics}
The implementation is carried out by the {\tt C++}  finite element library {\tt deal.II} \cite{dealii82}. {The Bi-conjugate gradient stabilized method \cite{bicgstab}\ is used as linear solver for SDLS and CFEM-SAAF-CLS while LS and CFEM-SAAF are solved using conjugate gradient method. Symmetric successive overrelaxation \cite{numerical}\ is used as preconditioner for all calculations with relaxation factor fixed at 1.4}. In all tests, we also include results from solving the globally conservative self-adjoint angular flux (SAAF) equation with CFEM as a comparison\cite{morel_saaf,yaqi_void}. In all problems, piecewise linear polynomial basis functions are used. {In the 1D examples, the angular quadrature is the Gauss quadrature. In the 2D test, the quadrature is a \red{Gauss-Chebyshev quadrature with azimuthal point number on each polar level specified in a way similar to level-symmetric quadrature}. See Ref.\ \cite{josh-quad}\ for further details}. 

\subsection{Reed's problem}
\begin{figure}[ht!]
	\begin{subfigure}{.5\textwidth}
		\begin{center}
			\hspace*{-2.5cm}\includegraphics[width=1.3\textwidth]{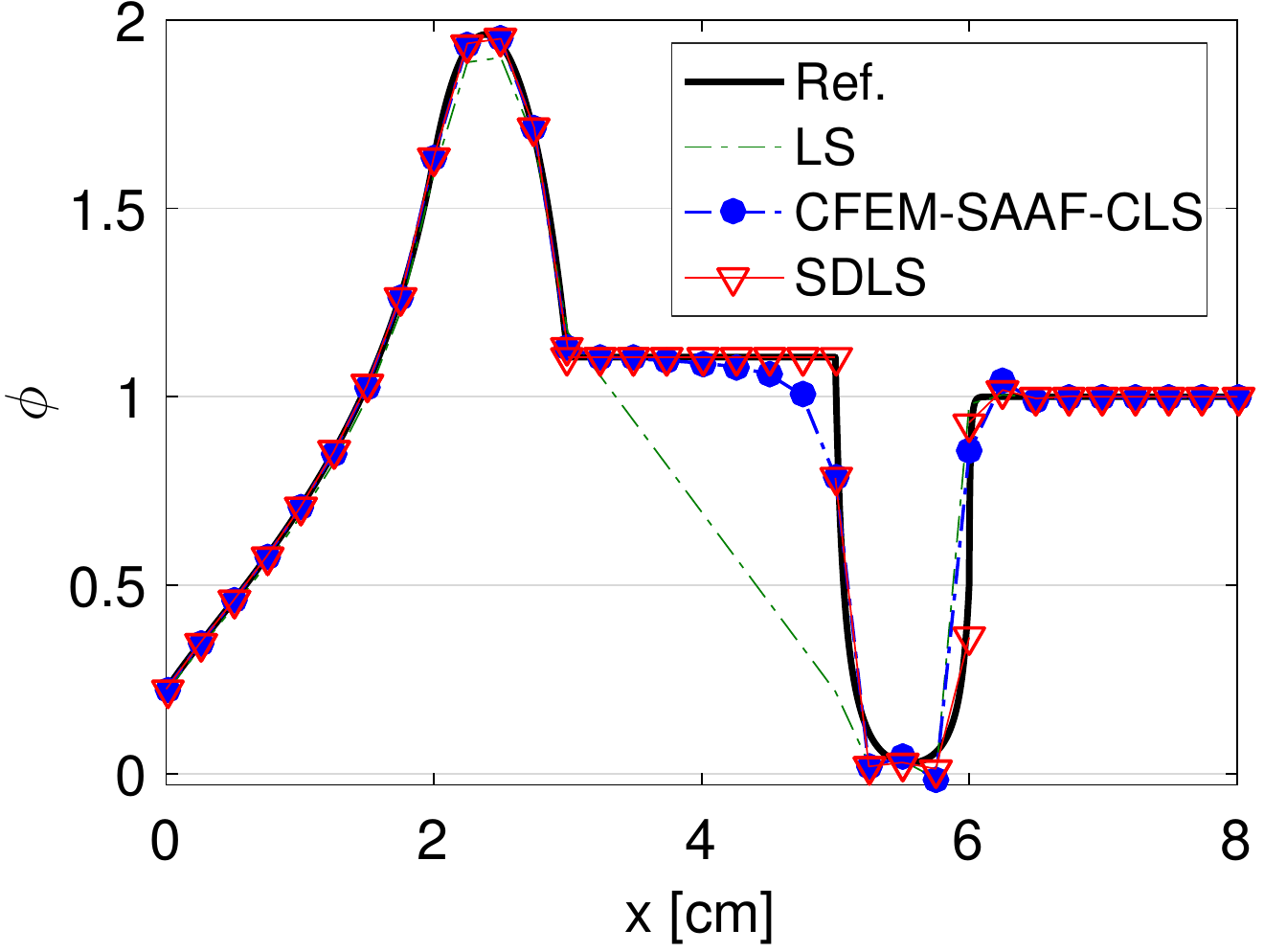}
			\caption[]{Reed's problem result comparison.}
			\label{f:reed-real}
		\end{center}
	\end{subfigure}
	~
	\begin{subfigure}{.5\textwidth}
		\begin{center}
			\includegraphics[width=1.3\textwidth]{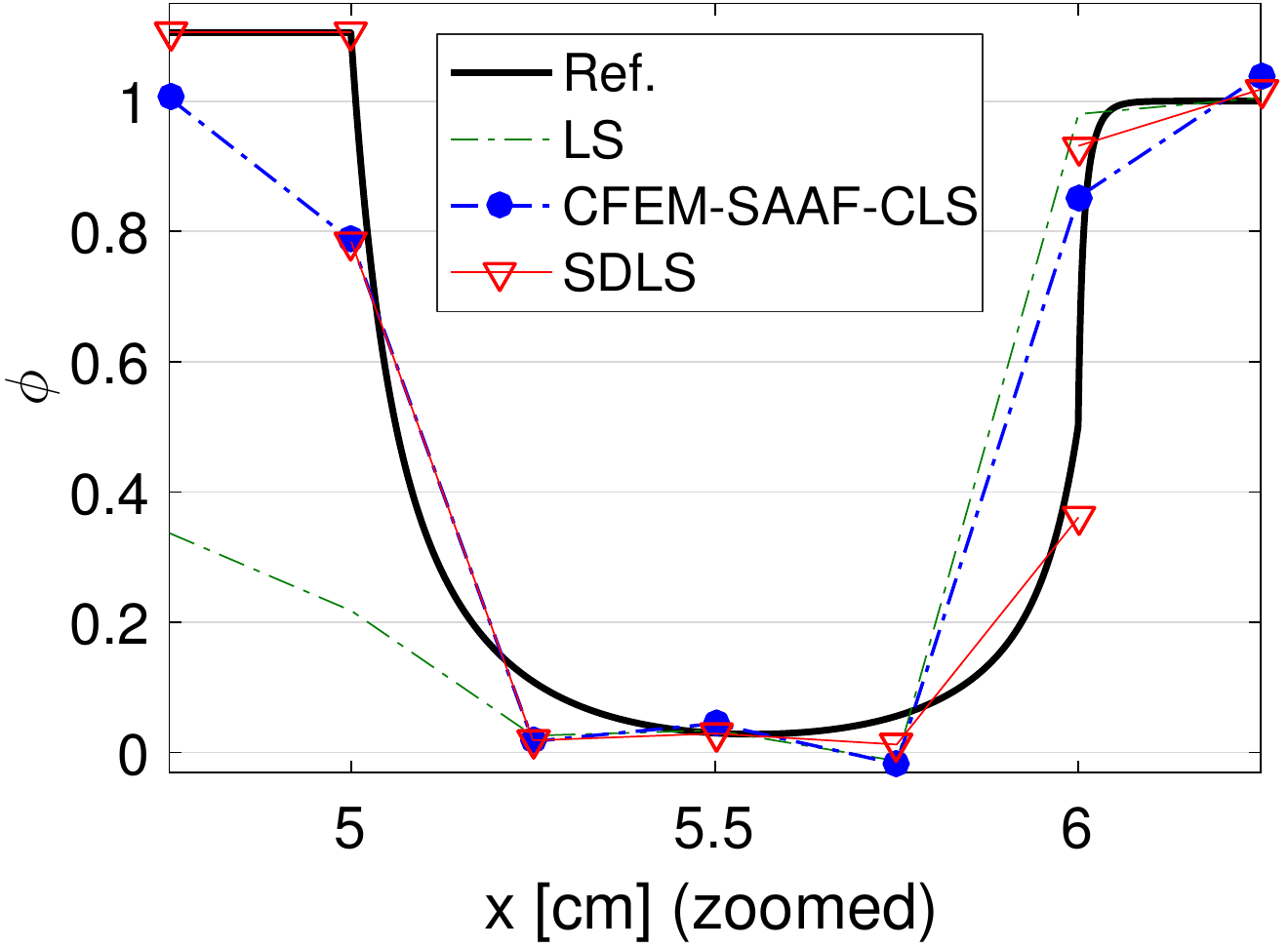}
			\caption[]{Zoomed results $x\in(4.75,~6.25)$\ cm.}
			\label{f:reed-real-zoomed}
		\end{center}
	\end{subfigure}
	\caption{Reed's problem results from different methods. 32\ cells are used.}
\end{figure}

{The first problem is  Reed's problem in 1D slab geometry \cite{reed_1971}\ designed to test spatial differencing accuracy and stability. The material properties are listed in Table\ \ref{tb:reed}.\ Specifically, a void region is set at $x\in(3,~5)$. Results using S$_8$\ in angle are presented in Figure\ \ref{f:reed-real}\ \red{with zoomed results for $x\in(4.75,6.25)$\ cm in Figure\ \ref{f:reed-real-zoomed}}.\ In this figure 32\ cells are used for the CFEM-LS (green line),\ CFEM-SAAF-CLS (blue line) and SDLS (red lines) calculations. For SDLS, the interfaces are set at $x=3,5,6$\ cm. The reference is provided by solving the first-order transport equation with {DFEM with linear discontinuous basis} using 2000\ cells. 
	
With the void treatment, the SDLS flux profile in $x\in(3,5)$\ is flat and the relative error is lower than $3\e{-5}$.\ By defining the relative balance as:}
\[{\bf B}_\mathrm{rel}=\frac{|{\bf B}|}{\displaystyle\intli{\ndo<0}\dome \intli{\pd}ds\ \absndo\psii{}+\intli{\mathcal{D}}dV\ Q},\]
{we found that the SDLS is globally conservative as ${\bf B}_\mathrm{rel}=5.56\e{-12}$.\ On the other hand, LS solution is distorted in void and the balance is poor (${\bf B}_\mathrm{rel}=3.73\e{-3}$). Meanwhile, we notice that CFEM-SAAF-CLS gives a better solution in the graph norm than LS in most regions of the problem. Yet, solution in void region is heavily affected by the thick absorber ($x\in(5,~6)$\ cm) as continuity is enforced at $x=5$\ cm.}

\begin{table}[ht!]
	\centering
	\caption{Material configuration for Reed's problem.}
	\label{tb:reed}
	\hspace*{0cm}\begin{tabular}{|c|c|c|c|c|c|}
		\hline
		x [cm] & (0,2) & (2,3) & (3,5) & (5,6)& (6,8)\\
		\hline
		$\st$ [cm$^{-1}$]& $1$ & $1$ & $0$ & $5$&$50$\\
		\hline
		$\sigs$ [cm$^{-1}$]& $0.9$ & $0.9$ & $0$ & $0$&$0$\\
		\hline
		$Q$& $0$ & $1$ & $0$ & $0$&$50$\\
		\hline
	\end{tabular}
\end{table}

\subsection{Two region absorption problem}
Another test problem is a 1D slab pure absorber problem firstly proposed by Zheng et al.\ \cite{zheng-inl}.\ There is a unit {isotropic} incident angular flux on left boundary of the slab. No source appears in the domain. In this problem \[\st= \begin{cases} 0.1\,  \text{cm}^{-1} &  x<1\, \text{cm} \\  10\, \text{cm}^{-1} & \text{ otherwise}\end{cases}.\]
\begin{figure}[ht!]
	\begin{subfigure}{.5\textwidth}
		\centering
		\hspace*{-1cm}\includegraphics[width=1.\linewidth]{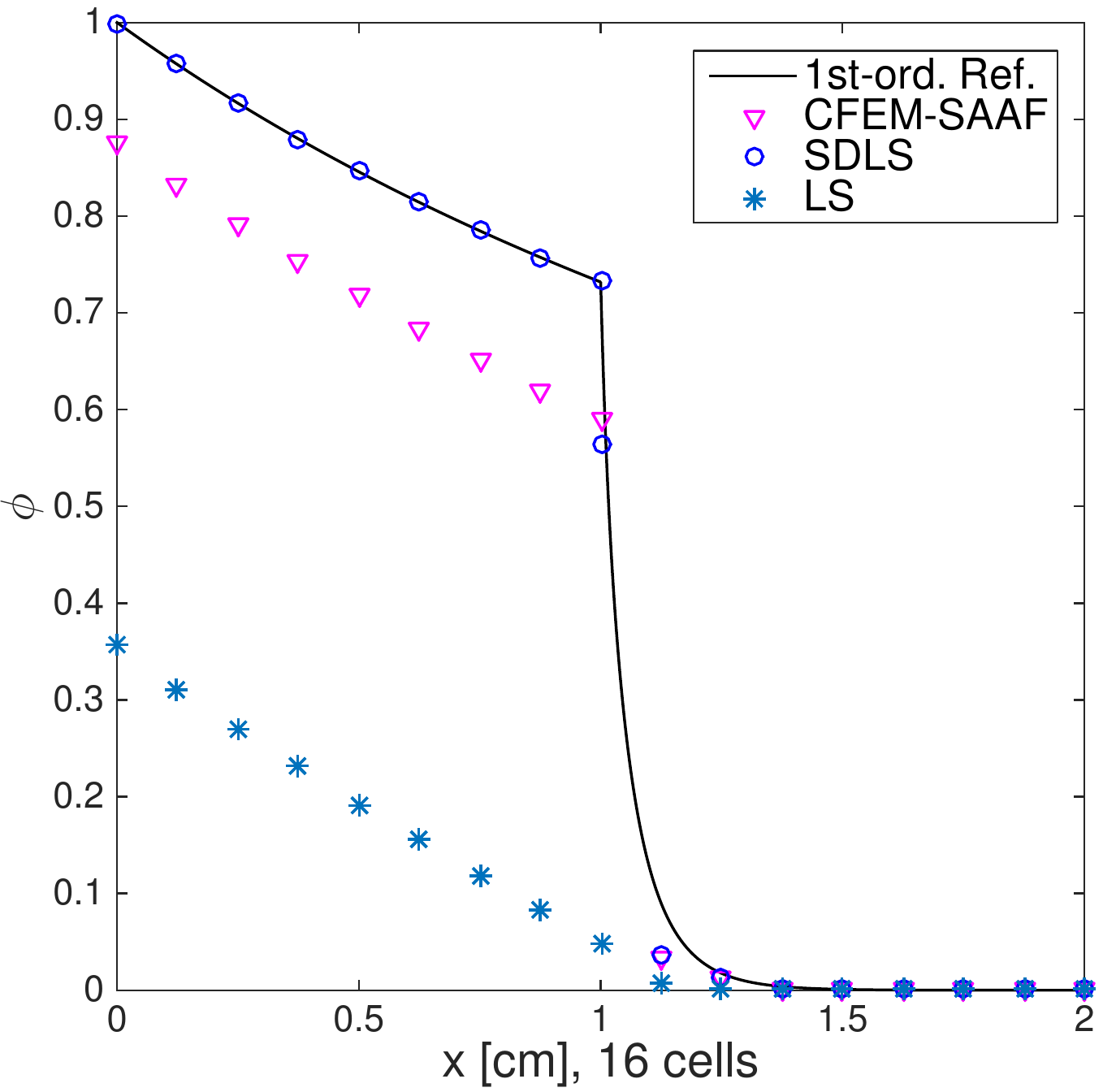}
		\caption{16 cells.}
		\label{f:2reg-16c}
	\end{subfigure}
	~
	\begin{subfigure}{.5\textwidth}
		\centering
		\includegraphics[width=1.\linewidth]{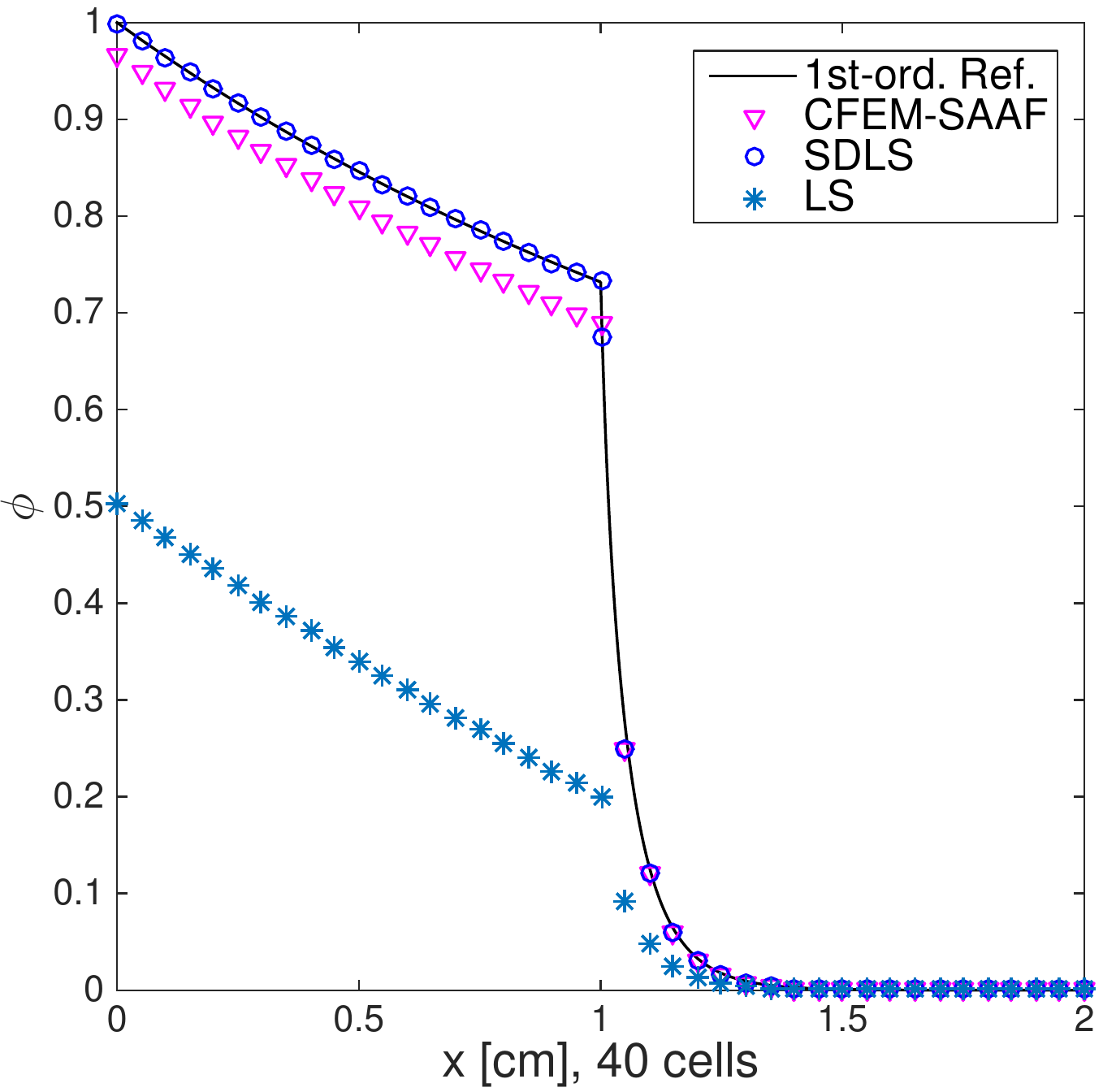}
		\caption{40 cells.}
		\label{f:2reg-40c}
	\end{subfigure}
	\caption{Two-region absorption results comparison with LS, SAAF and SDLS.}
	\label{mlfls}
\end{figure}

\begin{figure}[ht!]
	\begin{center}
		\includegraphics[width=.7\textwidth]{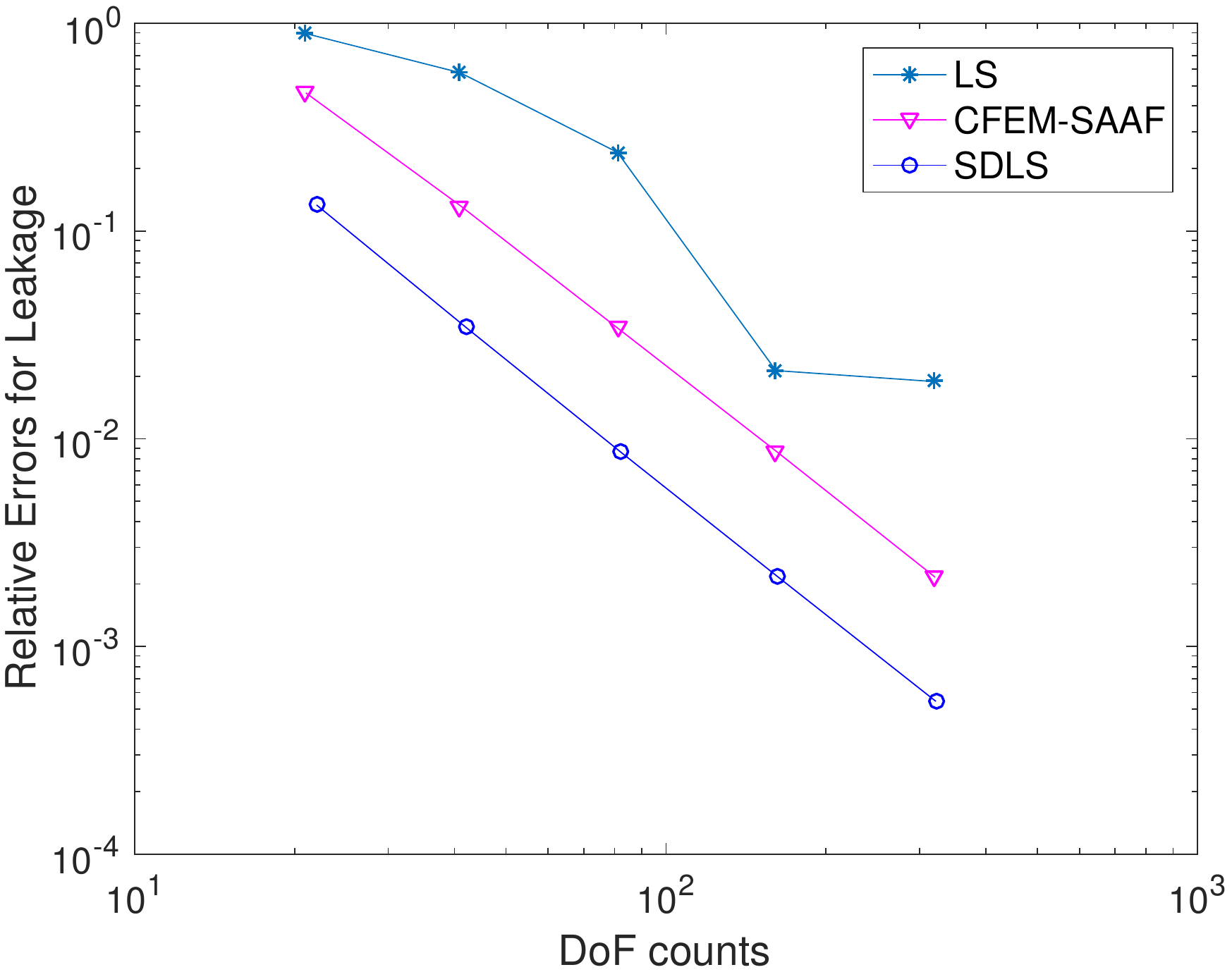}
		\caption[]{Leakage errors vs. DoF counts per direction at the right boundary.}
		\label{f:2reg-leak-error}
	\end{center}
\end{figure}

The results in Figure \ref{mlfls} compare coarse solutions using different methods. The reference solution was computed with first order S$_8$\ using diamond difference using $2\times10^4$\ spatial cells. The LS and SAAF solutions in the thin region ($x<1$\ cm) are affected by the presence of the thick region ($x>1$\ cm), even though that region is downstream of the incoming boundary source.  Increasing the cell count does improve these solutions, but there is still significant discrepancy with the reference solution. In both cases, the SDLS solution captures the behavior of the reference solution. In Figure \ref{f:2reg-leak-error} we see that both LS, SDLS and SAAF converge to the exact solution at second-order, with the convergence for LS being a bit more erratic.

Table \ref{tb:1}\ presents the relative global balances for different methods. As expected, LS presents large balance errors, even when the mesh is refined. On the other hand, CFEM-SAAF and SDLS give round-off balance to the iterative solver tolerance as expected. 

\begin{table}[ht!]
\centering
\caption{Relative global balances with different methods \red{(absolute values)}.}
\label{tb:1}
\hspace*{-1cm}\begin{tabular}{|c|c|c|c|c|c|}
\hline
Number of Cells & 20 & 40 & 80 & 160& 320\\
\hline
LS& $8.439\e{-1}$ & $6.426\e{-1}$ & $3.564\e{-1}$ & $3.430\e{-1}$&$4.948\e{-2}$\\
\hline
CFEM-SAAF&$5.899\e{-14}$&$1.786\e{-13}$&$2.615\e{-13}$  &$1.274\e{-12}$&$5.599\e{-12}$\\
\hline
SDLS&$2.148\e{-13}$&$7.668\e{-13}$&$1.492\e{-12}$  &$2.090\e{-11}$&$2.704\e{-12}$\\
\hline
\end{tabular}

\end{table}

%

\subsection{Thin-thick $k$-eigenvalue problem}
\begin{figure}[ht!]
	\begin{center}
		\includegraphics[width=.7\textwidth]{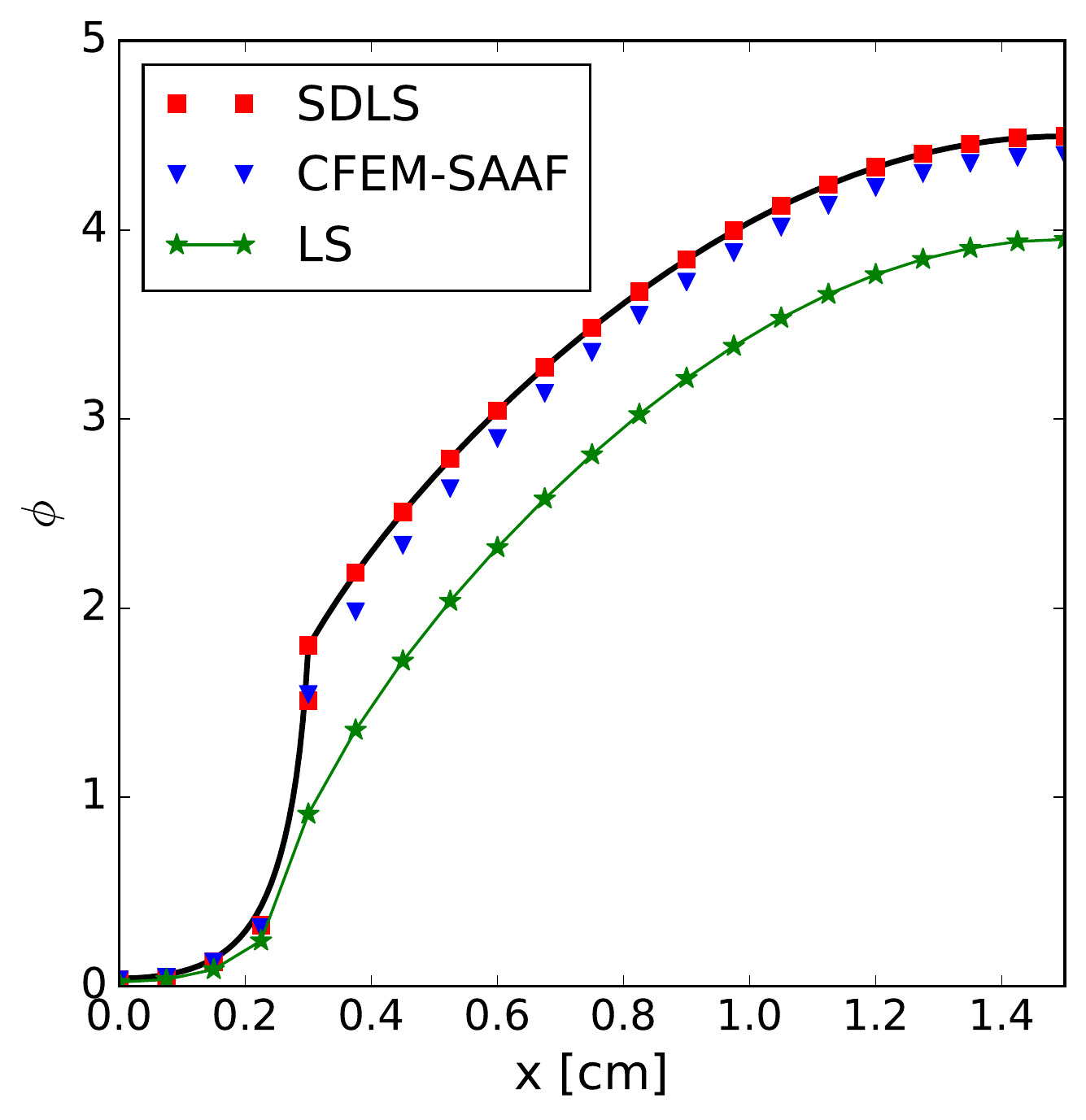}
		\caption[]{Scalar flux from different methods.}
		\label{f:eigen-flx}
	\end{center}
\end{figure}

{A one-group $k$-eigenvalue problem is also tested in 1D slab geometry. An absorber region in $x\in(0,0.3)$\ cm is set adjacent to a multiplying region in $x\in(0.3,1.5)$\ cm. Material properties are presented in Table\ \ref{tb:eigen}.\ Reflective BCs are imposed on both sides of the slab. The configuration is to mimic the impact from setting a control rod near fuel. 
	
A reference is provided by CFEM-SAAF using 20480\ spatial cells. With the presence of the strong absorber, the scalar flux has a large gradient near $x=0.3$\ cm. Figure\ \ref{f:eigen-flx}\ compares the results from different methods using 20\ spatial cells. In this case, LS (\red{green starred line}) presents noticeable undershooting near the \red{subdomain} interface. Though CFEM-SAAF (blue triangles) is an improvement, it is still away from the reference in fuel region. SDLS (\red{red squares}), in contrast, agrees well with reference solution except at the discontinuity introduced on the material interface set at $x=0.3$\ cm.}

\begin{table}[ht!]
	\centering
	\caption{Material configuration for $k$-eigenvalue problem.}
	\label{tb:eigen}
	\hspace*{0cm}\begin{tabular}{|c|c|c|}
		\hline
		x [cm] & (0,0.5) & (0.5,1.5)\\
		\hline
		$\st$ [cm$^{-1}$]& $5$ & $1$\\
		\hline
		$\sigs$ [cm$^{-1}$]& $0$ & $0.99$\\
		\hline
		$\nu\sigma_\mathrm{f}$& $0$ & $0.25$\\
		\hline
	\end{tabular}
\end{table}

\begin{figure}[ht!]
	\begin{subfigure}{.5\textwidth}
		\centering
		\hspace*{-2cm}\includegraphics[width=1.1\linewidth]{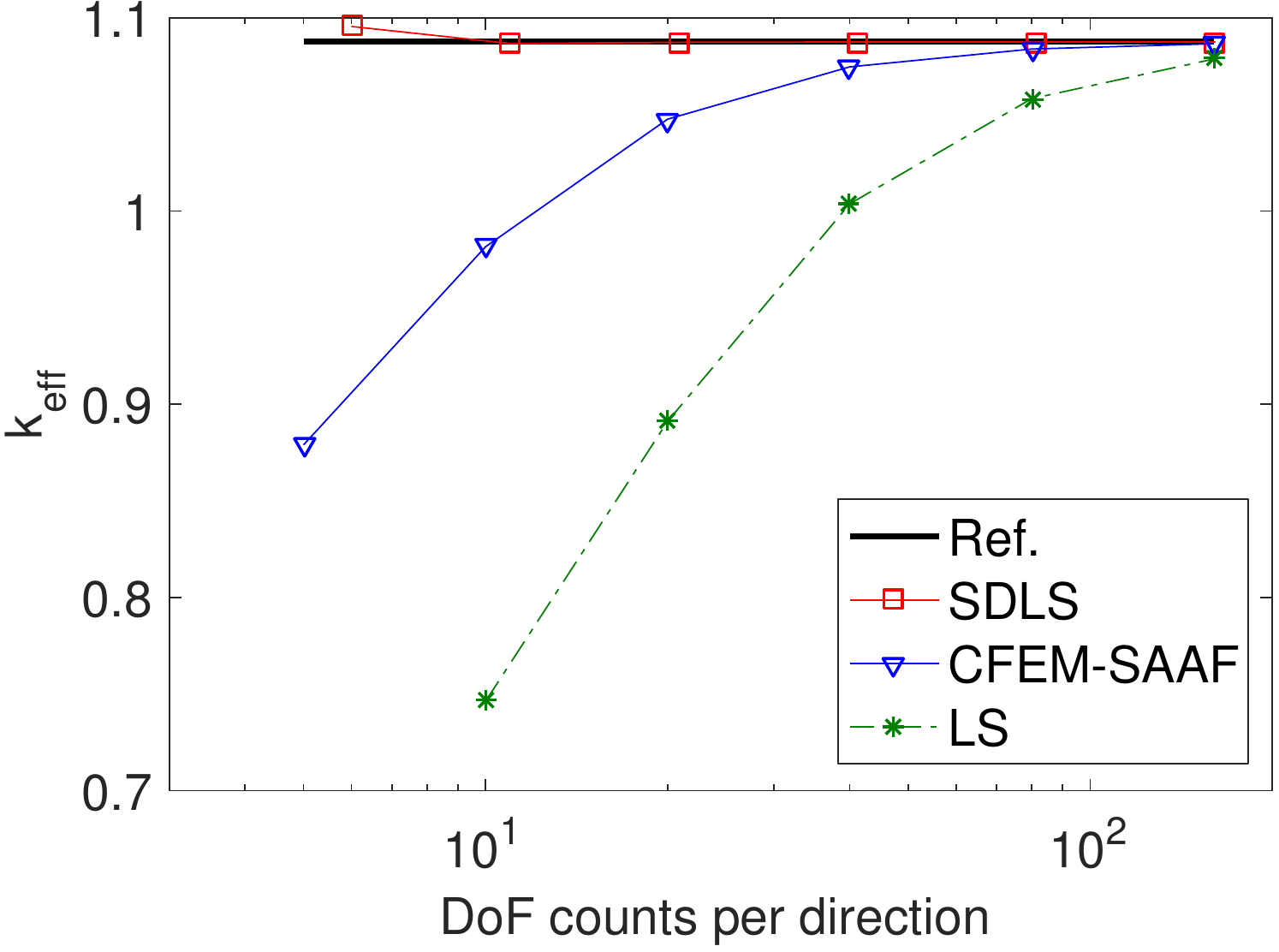}
		\caption{k$_\mathrm{eff}$\ values from different methods.}
		\label{f:k}
	\end{subfigure}
	~
	\begin{subfigure}{.5\textwidth}
		\centering
		\hspace*{-0cm}\includegraphics[width=1.1\linewidth]{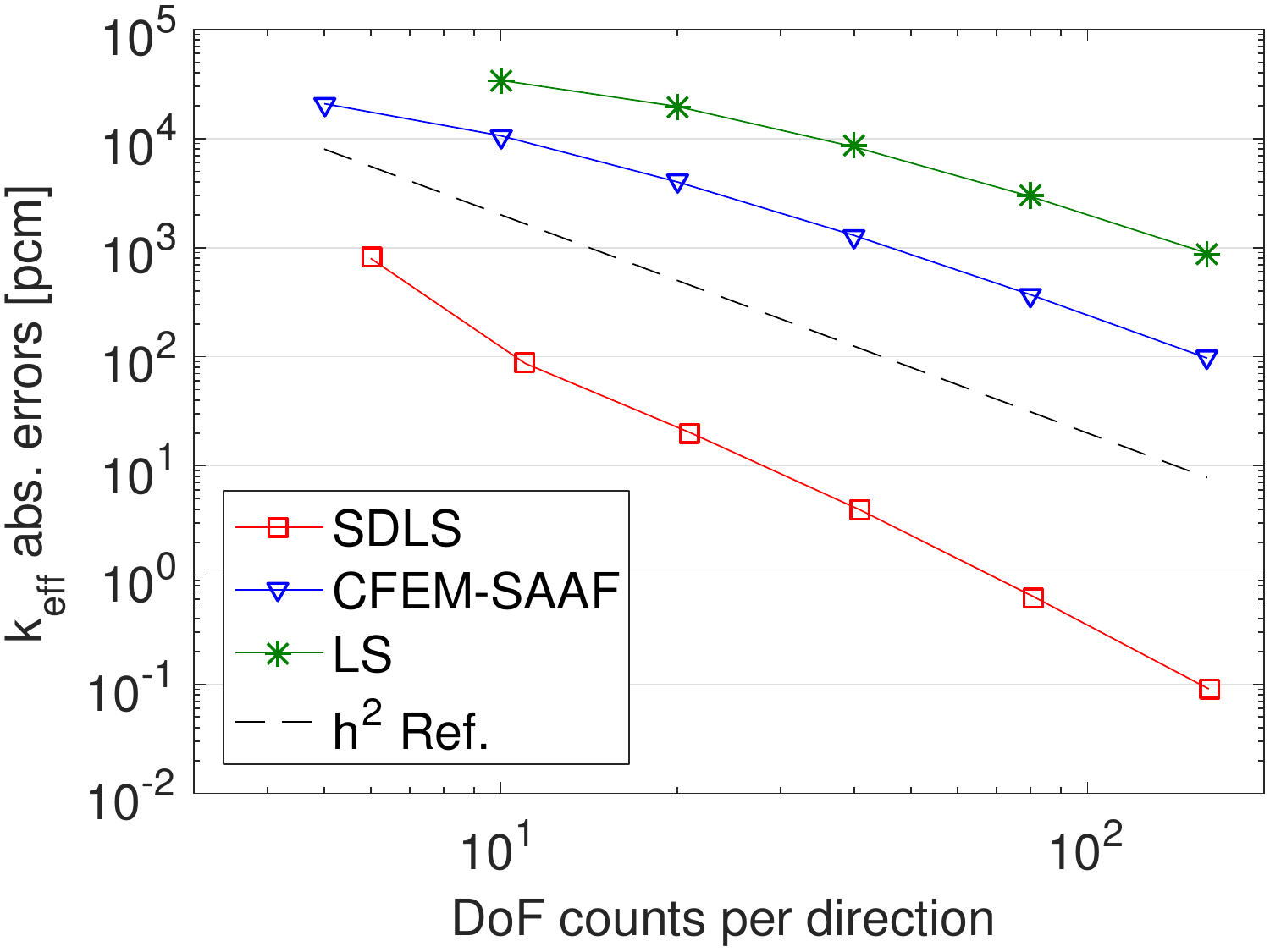}
		\caption{Absolute errors in pcm.}
		\label{f:k-err}
	\end{subfigure}
	\caption{k-eigenvalue results.}
	\label{f:eigen}
\end{figure}

{We also examine the $k_\mm{eff}$\ in Figure\ \ref{f:eigen}.\ We observe that SDLS converges to the reference $k_\mm{eff}$\ in the graph norm using only 10\ spatial cells in Figure\ \ref{f:k}.\ Yet, CFEM-SAAF slowly converges to the reference and LS still has a large error with 160\ cells. Figure\ \ref{f:k-err}\ shows $k_\mm{eff}$\ error change with respect to DoF counts and illustrates the benefit of adding the subdomain interface discontinuity. With 5\ cells (the points with smallest DoF counts), SDLS has over an order lower of $k_\mm{eff}$\ error than CFEM-SAAF. When refining the spatial mesh, SDLS shows roughly second-order convergence and has an error three orders of magnitude lower than CFEM-SAAF with 160 cells (the points with largest DoF counts) \red{although both have the global conservations}.}

\subsection{One group iron-water problem}
\begin{figure}[ht!]
	\begin{subfigure}{.5\textwidth}
		\centering
		\hspace*{-2cm}\includegraphics[width=.95\linewidth]{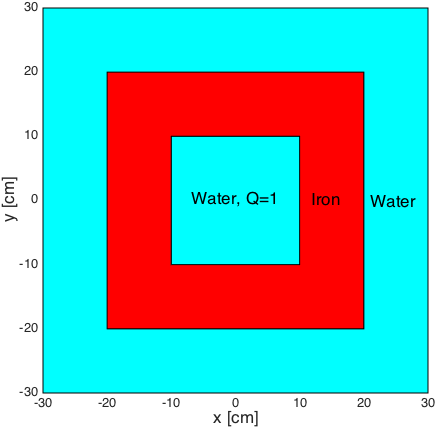}
		\caption{Problem configuration.}
		\label{f:config}
	\end{subfigure}
	~
	\begin{subfigure}{.5\textwidth}
		\centering
		\hspace*{-0cm}\includegraphics[width=1.1\linewidth]{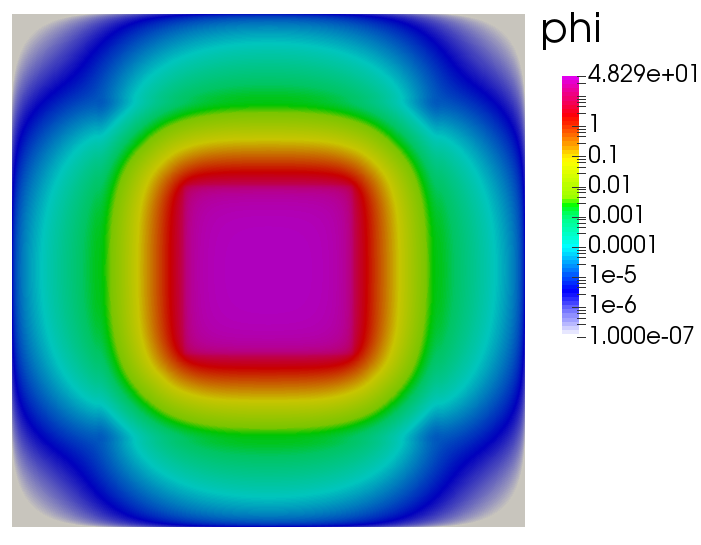}
		\caption{Scalar flux distribution from reference.}
		\label{f:pcolor}
	\end{subfigure}
	\caption{Results from the iron water problem.}
	\label{f:iron-water}
\end{figure}

\begin{figure}[ht!]
	\begin{subfigure}{.5\textwidth}
		\centering
		\hspace*{-2cm}\includegraphics[width=.95\linewidth]{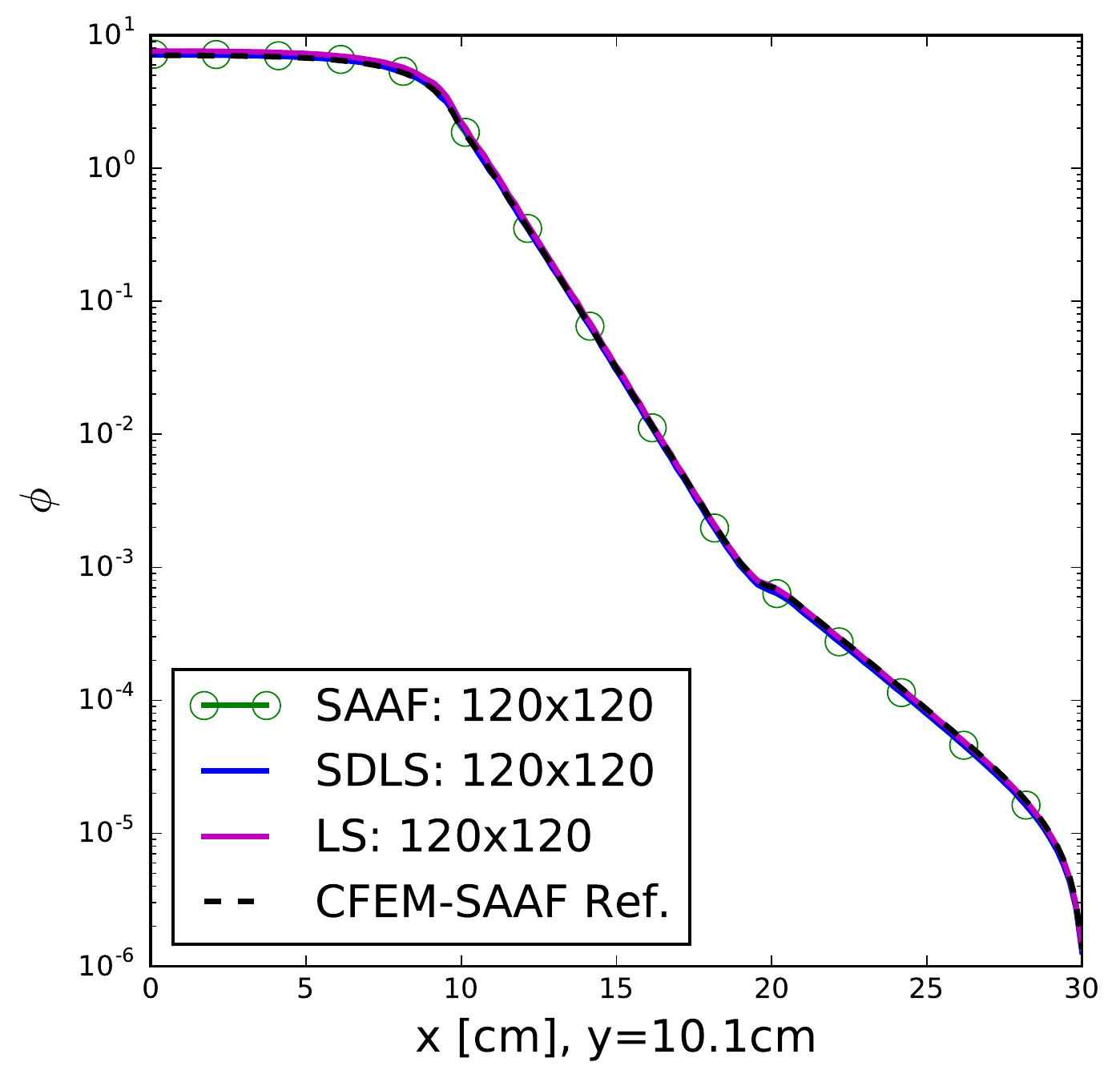}
		\caption{Line-out plot for $x=10.1$\ cm.}
		\label{f:line}
	\end{subfigure}
	~
	\begin{subfigure}{.5\textwidth}
		\centering
		\hspace*{-0cm}\includegraphics[width=1.4\linewidth]{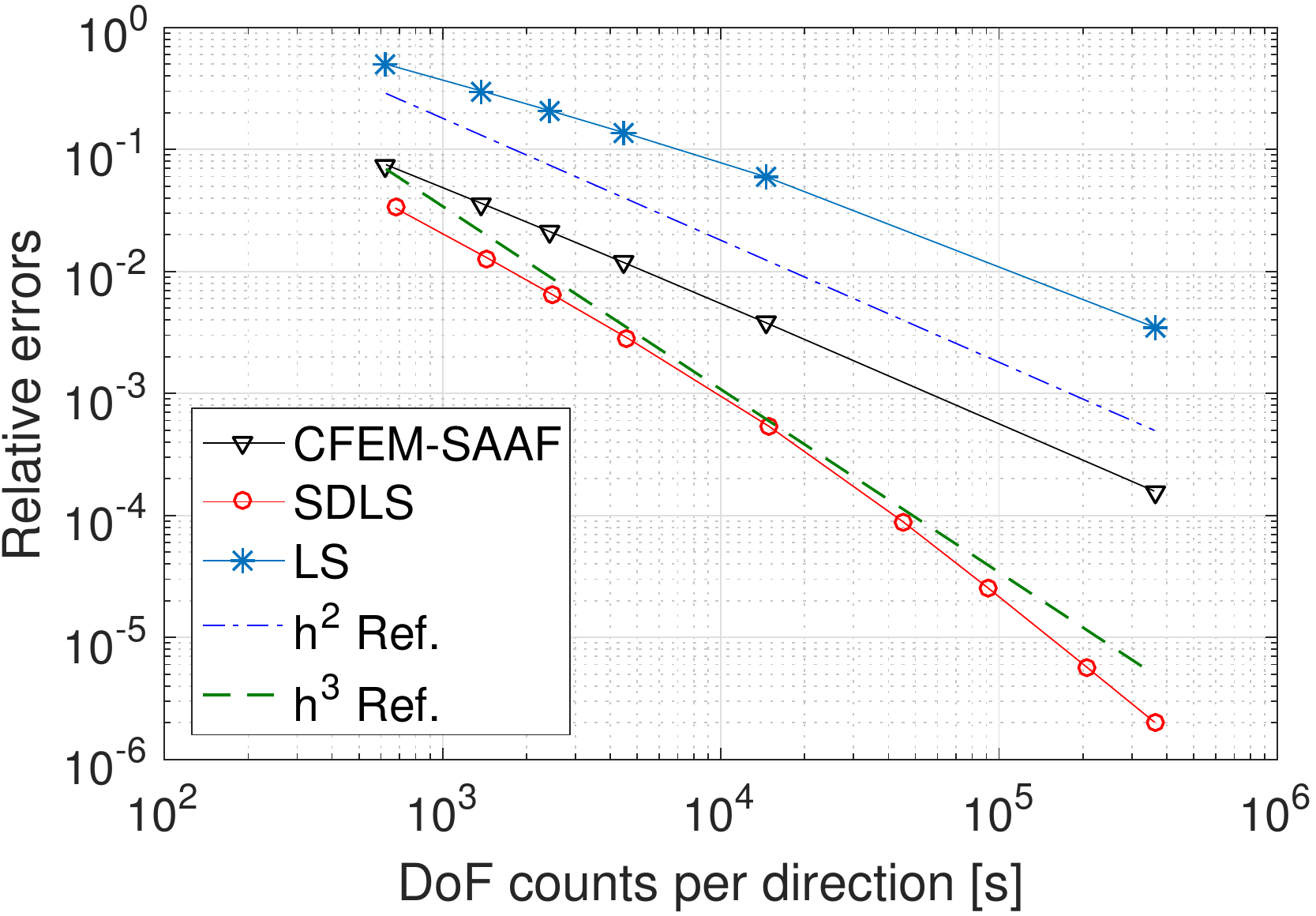}
		\caption{Relative iron absorption errors.}
		\label{f:e_abs}
	\end{subfigure}
	\caption{Results from the iron water problem.}
	\label{f:iw}
\end{figure}

The last test is a modified one-group iron-water shielding problem \cite{adams_iron_water}\ used to test accuracy of numerical schemes in relatively thick materials (see the configuration in Figure\ \ref{f:config}). Material properties listed in Table\ \ref{tb:iw}\ are from the thermal group data.\ S$_4$\ is used in the angular discretization. The reference is {SDLS} using $1200 \times 1200$ cells (see the scalar flux in Figure\ \ref{f:pcolor}).\ The scalar flux in the domain is rather smooth due to the scatterings. As a result, with moderately fine mesh (120$\times$120\ cells), we see SDLS graphically agrees with CFEM-SAAF and LS in the line-out illustrated in Figure\ \ref{f:line}.\ We further examine the absorption rate errors in iron. As shown in Figure\ \ref{f:e_abs},\ LS and CFEM-SAAF presents similar spatial convergence rates. However, LS presents lower accuracy than CFEM-SAAF with the presence of material interface between iron and water. On the other hand, by setting two interfaces between iron and water and introducing extra DoFs on the \red{subdomain} interfaces, SDLS converges to the reference solution much earlier than the other methods.

\begin{table}[ht!]
	\centering
	\caption{Material cross sections in one-group iron-water test.}
	\label{tb:iw}
	\hspace*{-1cm}\begin{tabular}{|c|c|c|}
		\hline
		Materials & $\st$\ [cm$^{-1}$] & $\sigs$\ [cm$^{-1}$]\\
		\hline
		Water& $3.2759$ & $3.2656$ \\
		\hline
        Iron&$1.1228$&$0.9328$\\
		\hline
	\end{tabular}
\end{table}

\begin{figure}[ht!]
	\centering
	\hspace*{-0cm}\includegraphics[width=.8\linewidth]{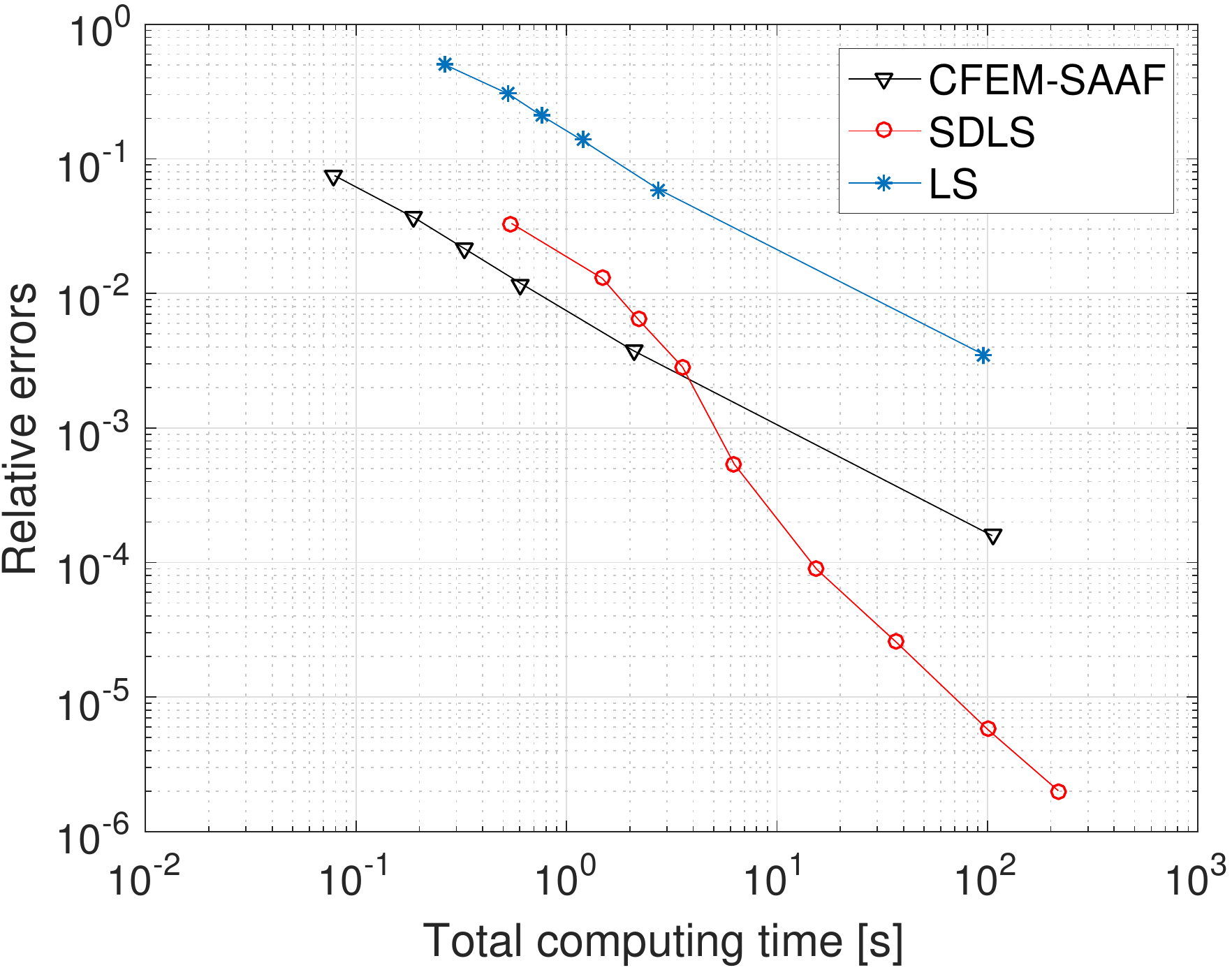}
	\caption{Relative errors of absorption rate in iron vs total computing time.}
	\label{f:e_abs-time}
\end{figure}

{In addition, we present a timing comparison of different methods in Figure\ \ref{f:e_abs-time}\ using the same error data employed in Figure\ \ref{f:e_abs}.\ Both CFEM-SAAF and SDLS are more accurate given a specific computing time. For a specific high error level ($>3\e{-3}$)\ obtained using coarse meshes,\ the SDLS is more time-consuming than CFEM-SAAF. However, for low error levels ($\red{<}3\e{-3}$)\ achieved  by refining the mesh, SDLS cost much less total computing time than CFEM-SAAF. On the other hand, when the total computing time increases, SDLS error drops faster than both CFEM-SAAF and LS. Overall, the linear solver presents reasonable efficiency for SDLS despite the fact that the system is non-SPD.}

\section{Concluding remarks and further discussions}\label{s:conc}
\subsection{Conclusions}
In this work, we proposed a subdomain discontinuous least-squares discretization method for solving neutral particle transport. It solves a least-squares problem {in} each region of the problem {assuming spatially uniform cross sections within each region} and couples the contiguous regions with discontinuous interface conditions. We demonstrated that our formulation preserves conservation in each subdomain and gives smaller numerical error than either SAAF or standard least squares methods. 

Though SDLS is angularly discretized with \sn\ method in this work, \pn\ would be applicable in the angular discretization as well. Therein, LS\pn\cite{zheng_invite,manteuffel_lspn_scaling,varin_lspn}\ is applied in each subdomain with Mark-type boundary condition used as interface condition \cite{sanchez-transport}.\ Further, since SDLS allows the discontinuity on the subdomain interface, different angular schemes, e.g. \sn\ and \pn\ can be used in different subdomains. {In fact, the angular coupling is desirable for SAAF in the code suite {\tt Rattlesnake}\ \cite{yaqi-rattlesnake}\ at Idaho National Laboratory. An initial implementation of the coupling has been developed in \cite{yaqi_physor16}\ through enforcing strong continuity of angular flux on coupling subdomain interfaces. And later, the interface discontinuity is applied to SAAF to allow an effectively improved coupling scheme reported in \cite{yaqi_invite,zheng-inl}\ and implemented in {\tt Rattlesnake}\ to improve the neutronics calculations.}

\subsection{Future recommendations}
{We restrict the method to the configuration that total cross section is uniform within subdomains. In cases total cross sections varies smoothly within subdomains, one could simply weight the subdomain functional with reciprocal total cross section while weighting the \red{subdomain} interface functional with $1$.\ The rationale is such that weak form resembles the conservative CFEM-SAAF in the corresponding subdomain with spatially varying cross sections (see Ref.\ \cite{zheng_dissertation}\ for demonstration).}

{The other interesting direction is to explore the effects of re-entrant curved \red{subdomain} interfaces, which is not considered in current study. The efficacy of the method and according solving techniques in these cases are not known and worthy of investigation.}
\section*{Acknowledgements}
W. Zheng is thankful to Dr. Bruno Turksin from Oak Ridge National Laboratory for the suggestions on multi-D implementation. Also, Zheng appreciates Dr. Yaqi Wang from Idaho National Laboratory for his suggestions to make the notation simple and clear. He wants to give appreciation to Hans Hammer from Texas A\&M Nuclear Engineering for running reference tests as well. \red{Finally, we wish to thank the anonymous reviewers with their invaluable advice to improve the paper to current shape}.

This project was funded by Department of Energy NEUP research grant from Battelle Energy Alliance, LLC- Idaho National Laboratory, Contract No: C12-00281.
\section*{References}

\end{document}